# Geometrical control of the magnetization direction in high aspect-ratio PdNi ferromagnetic nano-electrodes

*J. J. Gonzalez-Pons, J. J. Henderson, and E. del Barco\**

Department of Physics, University of Central Florida, 4000 Central Florida Blvd., Orlando, Florida 32816-2385

*B. Ozyilmaz*

Department of Physics, National University of Singapore, 2 Science Drive 3, Singapore 117542

*E-mail: delbarco@physics.ucf.edu

**Abstract:**

We present a study of electron-beam evaporated $Pd_{0.4}Ni_{0.6}$ alloy thin films by means of ferromagnetic resonance measurements on extended films of varying thickness and anisotropic magnetoresistance measurements of lithographically patterned high aspect-ratio ferromagnetic electrodes, respectively. The results reveal that the direction of the magnetization strongly depends on the electrode lateral dimensions, transitioning from in-plane magnetization for extended films to out-of-the-plane magnetization for electrode widths below 2-3 microns, reaching ~58 degrees off-plane for 100 nm-wide nanoelectrodes.

Recently $Pd_{1-x}Ni_x$ alloy has attracted considerable attention as ferromagnetic electrodes in carbon based lateral spin valves. Its excellent wetting properties on carbon nanotubes (CNT) [1-3], leads to low ohmic contacts (transparent), while its room temperature ferromagnetic behavior [4] provides a means for spin injection. Surprisingly in the case of CNTs a tunneling barrier between the PdNi and the CNT itself seems not to be necessary for spin injection, making PdNi alloys the ideal material for magnetic electrodes in low dimensional carbon-based electronic devices. Key to these experiments is anisotropy, since the successful demonstration of spin injection is based on the observation of a giant magnetoresistance (GMR) effect [5], which is determined by the relative orientation between the magnetization of the two ferromagnetic electrodes in a spin-valve device. In particular, it has been shown that in all-metallic spin-valves in which the magnetization vectors of both ferromagnets are not collinear ($0 < \theta < 180$), the current-induced switching of the magnetization state of the device can be obtained in much shorter times, making them highly efficient systems for information processing [6]. In carbon-based spin-valve devices the planar arrangement of the system complicates the realization of non-collinear magnetizations, and the ability to control the magnetization direction with respect to the plane of the electrode becomes essential. However little is known about the magnetic properties of such ferromagnetic alloys when lithographically patterned into narrow electrodes with large aspect ratios. A detailed understanding of the magnetic characteristics of this material is therefore crucial to both understand the switching characteristics of such a device and optimize the electrode dimension and configuration when used as ferromagnetic spin injectors.

In this letter we present a detailed study of the effect of the geometry of electron-beam patterned $Pd_{0.4}Ni_{0.6}$ ferromagnetic thin film structures on the equilibrium direction of the magnetization with respect to the film plane. For extended films, room temperature ferromagnetic resonance (FMR) measurements show that the magnetization remains in the film plane preferentially for 4-80 nm-thick films. A substantial out-of-the-plane uniaxial anisotropy which tends to pull the magnetization off-plane competes with the demagnetizing fields which set the magnetization in-plane. For laterally constrained large aspect ratio $Pd_{0.4}Ni_{0.6}$ thin film electrodes, anisotropic magneto-resistance (AMR) transport measurements show that the magnetization cants out of the film plane for electrode widths below ~2-3 µm, reaching an angle with respect to the film plane of ~58° for electrode widths down to ~100 nm.

$Pd_{0.4}Ni_{0.6}$ alloy extended films were fabricated by electron beam evaporation of the bulk material on $Si/SiO_2$ wafer in a UHV system with a base pressure of $7\times10^{-7}$ Torr. In addition, 25 nm-thick $Pd_{0.4}Ni_{0.6}$ films were patterned in the shape of high aspect ratio electrodes of length 20 µm and various widths (100 nm – 10 µm). In a second lithographic step the PdNi electrodes were contacted with standard Ti/Au electrodes (5/50nm) necessary for transport measurements.

FMR measurements were carried out at room temperature with a high-frequency broadband (1-50 GHz) micro-coplanar-waveguide (µ-CPW) [7] using the flip-chip method [8-10]. A 1.5 Tesla rotateable electromagnet was employed to vary the applied field direction from the in-plane ($\phi = 0°$) to normal to the film plane ($\phi = 90°$) directions. The change of the resonance field value of the 15 GHz FMR absorption peak as a function of the relative angle between the external magnetic field and a 20 nm-thick NiPd film is shown in Fig. 1. Inset in the figure there is a geometrical representation of the applied magnetic field vector and the angle, $\phi$, that was rotated through. Also inset in Fig. 1 there are two absorption curves corresponding to two orientations of the field, $\phi = 0$ (in-plane) and 90° (out-of-the-plane). Similar behavior was observed for all the studied films. The data are typical of polycrystalline ferromagnetic films with in-plane magnetization [9,11]. The resonant field value is lowest at 0 or 180 degrees, when the applied field is in the same plane as the magnetization vector.

The magnetic energy of a ferromagnetic film in the presence of a magnetic field, $H$, applied at an angle, $\phi$, with respect to the plane of the film is given by [9,11]

$$E = -M_s H(\cos\phi\cos\phi_m + \sin\phi\sin\phi_m) + 2\pi M_s^2 \sin^2\phi_m - (K_1 + 2K_2)\sin^2\phi_m + K_2 \sin^4\phi_m \quad (1)$$

where $M_s$ is the saturation magnetization, $\phi_m$ is the angle between the magnetization and the film plane and $K_1$ and $K_2$ are the first and second order out of-the-plane uniaxial anisotropy constants, respectively. The first term represents the Zeeman energy associated to the coupling of the magnetization and the external field, which tries to align both along the same direction. The second term is the magnetostatic energy, which forces the magnetization into the plane. The last terms represent the out-of-the-plane uniaxial anisotropy, which minimizes the energy in a direction normal to the plane. The final orientation of the magnetization of the ferromagnetic film results from a competition between these three energies. The data in Fig. 1 can be fitted using the equation of condition of resonance given by the Smit and Beljers formula [12], $\omega = \gamma\sqrt{H_1 H_2}$, where $\gamma = g\mu_B/\hbar$ is the gyromagnetic ratio, and $H_1$ and $H_2$ depend on $H$, $\phi_m$, $\phi$, $M_s$, $K_1$ and $K_2$ (see Refs. [9] and [11] for the exact expression). A good fitting of the data in Fig. 1 (continuous line) is obtained for $M_s = 290$ emu/cm$^3$ (according to a 60%-Ni composition with magnetization density $M_s^{Ni} = 485$ emu/cm$^3$), $g = 2.22$, $K_1 = 2.26\times10^{-5}$ erg/cm$^3$ and $K_2 = 0.22\times10^{-5}$ erg/cm$^3$.

When the field is applied at $\phi = 0$ or $90°$, the resonance condition reduces to:

$$\begin{aligned}\left(\frac{\omega}{\gamma}\right)^2_{//} &= H_{res}\left(H_{res}+[4\pi M_{eff}]_{//}\right)\\ \left(\frac{\omega}{\gamma}\right)_{\perp} &= H_{res}-[4\pi M_{eff}]_{\perp}\end{aligned} \qquad (2)$$

where $[4\pi M_{eff}]_{//} = 4\pi M_s - 2K_1/M_s - 4K_2/M_s$, and $[4\pi M_{eff}]_{\perp} = 4\pi M_s - 2K_1/M_s$. According to this, the difference between the frequency dependences of the FMR peak in the parallel and perpendicular configurations (see Eq. 2) enables an independent determination of the anisotropy parameters. The longitudinal and perpendicular frequency behaviors of the FMR peak of a 20 nm thick film are shown in Fig. 2. The slope of the curves allows the determination of the gyromagnetic ratio, hence the Landé $g$-factor [13], while the intersept with the $x$-axis gives the effective in-plane and out of plane magnetizations, $[(4\pi M_{eff})]_{//}$ and $[(4\pi M_{eff})]_{\perp}$. The difference observed between the two effective demagnetization fields is indicative of the presence of a second order anisotropy term, which can be deduced from ($[4\pi M_{eff}]_{//} - [4\pi M_{eff}]_{\perp} = 4K_2/M_s$). The thickness dependence of the first and second order anisotropies, $K_1$ and $K_2$, extracted by this method is shown in inset to Fig. 2. The results indicate that the out-of plane ($K_1, K_2 > 0$) easy-axis anisotropy first decreases with increasing film thickness, for thickness values below 10nm, and reaches a minimum ($K_1 = 1.3 \times 10^{-5}$ erg/cm$^3$) at $t = 10$ nm. $K_1$ then increases up to $2.9 \times 10^{-5}$ erg/cm$^3$ for a film thickness of 25 nm, above which it becomes practically thickness independent (results obtained up to 80 nm, not shown here, support this evidence). The magnetic quality factor of the thin film is defined as the ratio between the anisotropy energy and the demagnetizing (magnetostatic) energy, $Q = (K_1+K_2)/2\pi M_s^2$. In the limit of $Q < 1$, the demagnetization field dominates over the out-of-the-plane anisotropy. For the thicknesses studied here we obtain $Q < 0.7$, in agreement with the in-plane alignment of the magnetization extracted from the angular dependence of the FMR field. In general, the demagnetizing field vector is determined by the shape of the sample and characterized by the demagnetizing factors $N_x$, $N_y$ and $N_z$, with $N_x + N_y + N_z = 4\pi$. In the case of an extended thin film, the only component of the demagnetizing field is normal to the plane (i.e. $N_z = 4\pi$ and $N_z = N_y = 0$), and will strongly oppose to the magnetization to tilt away from the film plane.

Anisotropic magneto-resistance (AMR) measurements were performed in laterally constrained $Pd_{0.4}Ni_{0.6}$ large aspect ratio electrodes (i.e. nanowires) of thickness $t = 25$ nm, length $l = 20$ μm and widths varying from 100 nm to 10 μm. An AFM image of one of the electrodes is shown in the inset to Fig. 3. The magneto-resistance varies as a function of the relative angle between the applied electrical current and the magnetization vector, $\vec{M}$. The largest (lowest) resistance is obtained when $\vec{M}$ and $\vec{I}$ are collinear (perpendicular) [14]. The magneto-resistance is proportional to $\cos^2(\theta)$ [15,16], where $\theta$ is the angle between the magnetization and the current, which in our case flows along the length of the nanowire. Yet, the direction of the magnetization is determined by the direction and strength of the applied magnetic field, the intrinsic uniaxial anisotropy and the demagnetizing field of the film, which will in turn depend on the geometry (width) of the nanowire.

The AMR curves obtained with the external dc field applied along three different directions relative to the electrode geometry are shown in Fig. 3 for a 25 nm-thick $Pd_{0.4}Ni_{0.6}$ electrode 250 nm wide and 20 μm long. The largest change in resistance, a 0.68 % increase of the AMR at saturation, is observed when the magnetic field is aligned along the axis of the electrode (L). On the contrary, when the magnetic field is oriented perpendicular to the film, only a 0.352 % decrease of the AMR is observed. In both cases, the saturation of the AMR response is achieved for fields over ~ 2 kG, in agreement with the value of the effective demagnetizing fields observed by FMR in extended films (Figs. 2a-b). However, when the field is oriented across the wire there is no appreciable change in AMR with field for this particular electrode width, for field values as high as 8 Tesla. This is indicative of a modified demagnetizing field vector imposed by the constrained lateral dimension of the structure, which prevents the magnetization to orient across the wire. For widths larger than 2-3 µm, the transverse AMR saturates at the same value than the perpendicular AMR, as will be discussed below.

One can extract information about the direction of the magnetization by looking at the absolute change in AMR for different field orientations. In the case of the electrode shown in Fig. 3, the fact of having the largest AMR change when the field is applied along the nanowire axis, $\Delta_L = 0.68$ %, is indicative of an initial magnetic configuration in which most of the spins were aligned away from this direction, contributing to a large increase of the device resistance when aligning with the current by the action of the longitudinal field (L). In addition, the smaller AMR change (decrease) with the field applied perpendicular to the film plane (P) is an indication of the fact that most of the spins were already close to this orientation before the field application. Since for this particular nanowire there is no change in magneto-resistance, $\Delta_T = 0$, when the field is applied transversally to the wire (T), we can assume that all the spins are located in the L-P plane. In this situation and assuming a rigid magnetization vector, the normalized AMR in the longitudinal direction, $\Delta_{LN} = \Delta_L/(\Delta_L+\Delta_P)$, can be understood as the projection of the magnetization along the P-axis. Consequently, one can estimate the angle between the magnetization and the film plane using $\phi_m = \sin^{-1}(\Delta_{LN})$, which for the 250 nm wire in Fig. 3 corresponds to $\phi_m = \sin^{-1}(0.787) \sim 52°$.

Fig. 4a shows the normalized longitudinal AMR change as a function of the electrode width, $w$. It can be observed how the longitudinal AMR change reaches its maximum value ($\Delta_{LN} \sim 0.85$) for a 100 nm nanowire. Following the previous arguments, this corresponds to an angle of 58 degrees between the magnetization and the film plane. $\Delta_{LN}$ gradually decreases with increasing widths until arriving to a final value of 0.5 for widths over 2-3 µm. Before going into a detailed discussion about the magnetic configuration of the widest electrodes it is important to note that for electrode widths over ~1 µm there starts to appear a change in the transverse AMR, $\Delta_T$, which eventually arrives to the same saturation value than the perpendicular AMR change, $\Delta_P$, for widths over 3 µm. This indicates that the transverse direction (T) ceases to be a hard anisotropy axis for large electrode widths and an applied magnetic field has the same effect on the magnetization for different orientations within the plane, as in the case of an extended film. Naturally, different orientations of the magnetization within the plane correspond to different AMR values, since the current is applied along the L-axis. Note that for electrode widths over 3 µm, the absolute value of the normalized AMR change at saturation is the same for the three orientations of the field with respect to the electrode axes ($\Delta_L = \Delta_P = \Delta_T = 0.5$). This indicates that the magnetization lies in the plane of the electrode without a preferential orientation.

    The results indicate that the magnetization of the electrodes gradually depart from the film plane for widths below 3 µm, reaching a maximum angle of 58º for the lowest electrode width studied here ($w = 100$ nm). Note that an out-of-the-plane equilibrium configuration of the magnetization has already been observed in thin 100 nm-wide $Pd_{0.6}Ni_{0.4}$ at low temperature [3]. For large electrode widths, the situation becomes equivalent to the one found by FMR measurements on extended thin films, indicating that the lateral constriction of the electrodes forces the magnetization to lie out of the plane, allowing a geometrical control of the magnetization direction of PdNi ferromagnetic electrodes. We deduce that the constrained lateral dimension of the electrode modifies the demagnetizing fields of the device allowing the uniaxial anisotropy of the PdNi film to overcome the magnetostatic energy, pulling the magnetization off

plane. Note that the small width of the electrodes generates another non-zero component of the demagnetizing vector, $N_x > 0$. Since, as discussed above, the sum of the three demagnetization factors must remain constant ($N_x + N_y + N_z = 4\pi$), this implies a reduction of the demagnetizing factor perpendicular to the film, $N_z < 4\pi$. Consequently, a weaker perpendicular demagnetizing field will not be sufficient to sustain the in-plane magnetization of the extended film in its competition with the out-of-the-plane uniaxial anisotropy — an effect that becomes more intense the thinner the electrode. The effect of the device geometry on the reconfiguration of the equilibrium magnetization direction has been studied in other single-ferromagnet nanoscale films. For example, in Co/Au nanodots a transition from in-plane (at room temperature) to out-of-the-plane (at low temperature) magnetization is observed to be modulated by the thickness of the ferromagnetic layer [17].

The possibility to control the orientation of the magnetization in ferromagnetic electrodes is of crucial importance for producing efficient spin-transfer devices. Our results show that in PdNi alloy ferromagnetic thin films this can be achieved by tuning the geometry of the electrode. Specifically, the magnetization of $Pd_{0.4}Ni_{0.6}$ electrodes can be engineered to transit from in-plane to out-of-the-plane (up to 58 degrees) by varying the electrode width from 3 µm down to 100 nm respectively. The demonstrated tunability of this material could be used to fabricate planar devices in which different electrodes show different magnetization directions, allowing to probe the physics of spin-transfer seen in all-metallic spin-valves now in carbon-based devices, which could eventually lead to novel applications in emerging nanotechnologies.

**ACKNOWLEDGEMENTS:**

We want to acknowledge fruitful discussions with Jean-Marc L. Beaujour and Werner Keune. J.C.G acknowledges support from UCF Undergraduate Research Initiative program. J.C.G, J.J.H and E.d.B acknowledge support from the US National Science Foundation (DMR0706183 and DMR0747587).

**REFERENCES:**

[1]     S. Sahoo, T. Kontos, C. Schonenberger and C. Surgers, *App. Phys. Letters* **86**, 112109 (2005).

[2]     S. Sahoo, T. Kontos, J. Furer, C. Hoffmann, M. Graber, A. Cottet and C. Schonenberger, *Nature Physics*, **1,** 99-102 (2005).

[3]     H. T. Man, I. J. W. Wever and A. F. Morpurgo, *Phys. Rev. B* **73**, 241401(R) (2006)

[4]     W. A. Ferrando *et al.*, *Phys. Rev. B* **5**, 4657 (1972).

[5]     M. N. Baibich *et al*. *Phys. Rev. Lett.* **61**, 2472 (1988); G. Binasch, P. Grünberg, F. Saurenbach and W. Zinn, *Phys. Rev. B* **39**, 4828 (1989).

[6]     A. D. Kent, B. Ozeymalz and E. del Barco, *Appl. Phys. Lett.* **84**, 3897 (2004).

[7]     W. Barry, I.E.E.E Trans. Micr. Theor. Techn. MTT 34, 80 (1996).

[8]      G. Counil *et al.*, *J. Appl. Phys.* **95**, 5646 (2004).

[9]     J-M. L. Beaujour, W. Chen, K. Krycka, C. –C. Kao, J. Z. Sun and A. D. Kent, *Eur. Phys. J. B* **59**, 475-483 (2007).

[10]   J.-M. L. Beaujour, W. Chen, A. D. Kent and J. Z. Sun, *J. Appl. Phys.* **99**, 08N503 (2006).

[11]   C. Chappert, K. L. Dang, P. Beauvillain, H. Hurdequint and D. Renard, *Phys. Rev. B* **34** (5), 3192 (1986).

[12]   S. V. Vonsovskii, *Ferromagnetic Resonance*, Pergamon, Oxford, (1996).

[13]   The average effective value of the *g*-factor is 2.3 for the thickness range studied. This value is larger than the Landé *g*-factor of bulk Ni ($g_{Ni}$ = 2.18), which may be attributed to the moment induced in Pd by the surrounding Ni.

[14]   J. W. Loram and K. A. Mirza, *J. Phys. F: Met. Phys.* **15** 2213 (1985).

[15]   T. R. McGuire and R. I. Potter, *IEEE Trans. Magn.* **11**, 1018 (1975).

[16]   L. Gridneva *et al.*, *Phys. Rev. B* **77**, 104425 (2008)
11

**FIGURE CAPTIONS**

**Figure 1:** Behavior of the FMR peak position of a 20nm-thick $Pd_{0.4}Ni_{0.6}$ film as a function of the angle of application of the magnetic field from the plane of the field upon application of 15GHz microwave excitation. The sketch at the bottom of the figure represents the experimental configuration of the sample, which is placed upside-down on top of the µ-CPW transmission line. The magnetic field direction is also indicated. The inset on the left shows the corresponding FMR absorption versus field for angles $\phi = 0$ and $90°$.

**Figure 2:** Frequency behavior of the FMR peak of a 20nm-thick $Pd_{0.4}Ni_{0.6}$ film in the parallel (left axis, solid data) and perpendicular (right axis, open data) configurations, respectively. The inset shows the first ($K_1$) and second ($K_2$) order out-of-the-plane uniaxial anisotropies as a function of the film thickness.

**Figure 3:** (Color online) Room temperature AMR response of a 25nm-thick, 250nm-wide, 10µm-long $Ni_{0.6}Pd_{0.4}$ electrode as a function of an external magnetic field applied in the directions indicated in the figure. The inset on the left shows an AFM image of the electrode.

**Figure 4:** a) Normalized longitudinal AMR change as a function of the electrode width. The open circles were taken on a second sample that was fabricated independently. b) Ratio between the transverse and perpendicular AMR changes as a function of the width. The sketch in the middle represents the orientation of the magnetization with respect to the film plane for different electrode widths.

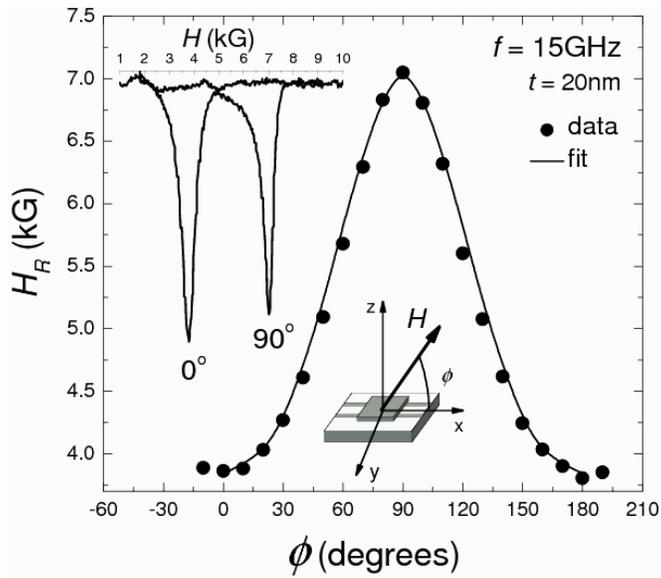

FIGURE 1

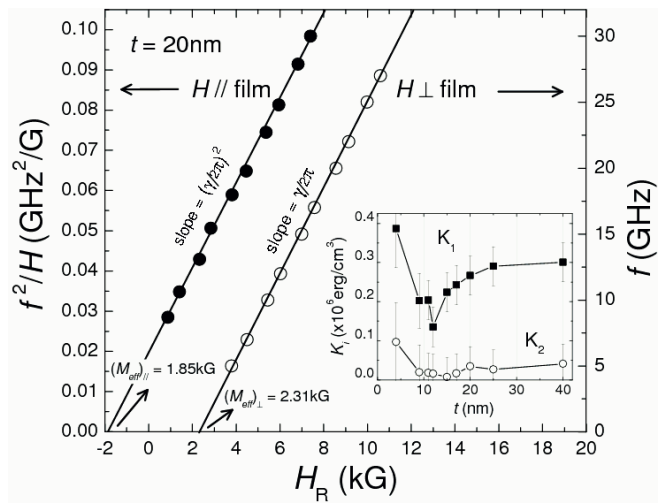

FIGURE 2

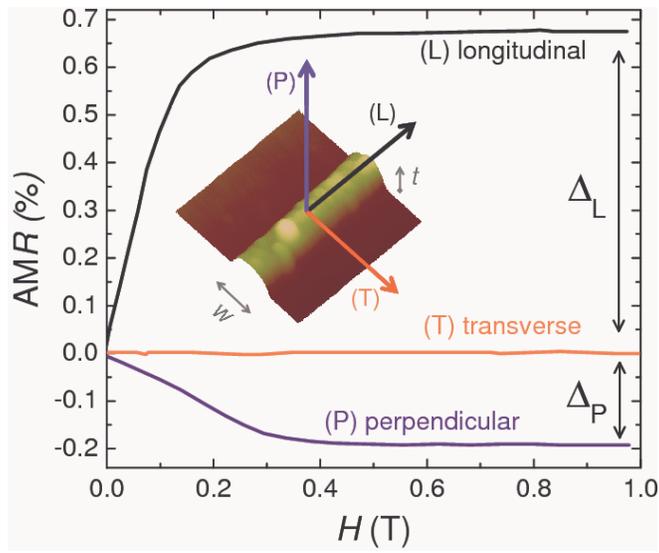

FIGURE 3

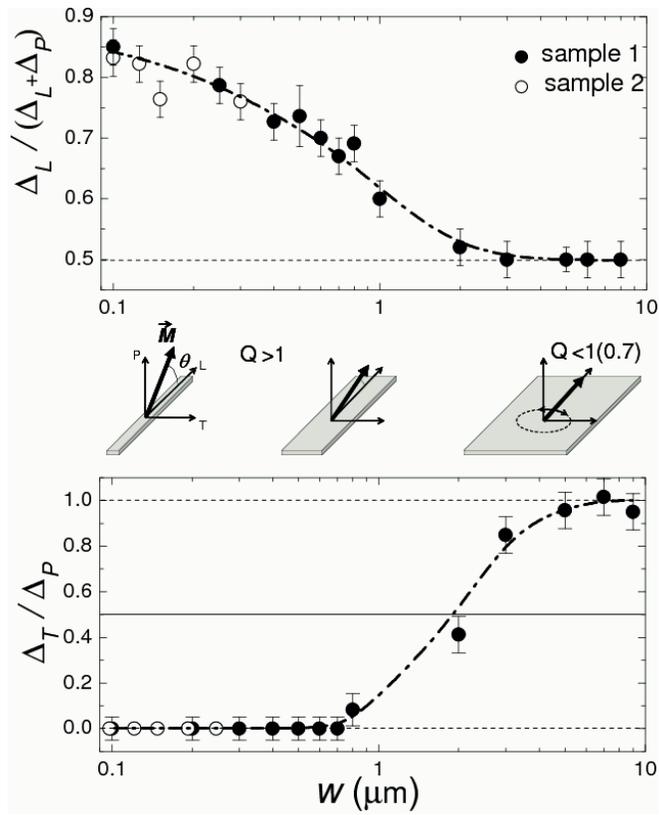

FIGURE 4